\documentclass[aps,prl,twocolumn,showpacs,groupedaddress]{revtex4-1}
\usepackage{graphicx}
\usepackage{epstopdf} 
\begin{document}

\title{Quantum Light from a Metal Nanoparticle}

% \affiliation command applies to all authors since the last
% \affiliation command. The \affiliation command should follow the
% other information
% \affiliation can be followed by \email, \homepage, \thanks as well.
\author{V.G.~Bordo}
\email{bordo@mci.sdu.dk}

\affiliation{NanoSYD, Mads Clausen Institute, Syddansk Universitet, Alsion 2, DK-6400 S{\o}nderborg, Denmark}

%Collaboration name if desired (requires use of superscriptaddress
%option in \documentclass). \noaffiliation is required (may also be
%used with the \author command).
%\collaboration can be followed by \email, \homepage, \thanks as well.
%\collaboration
%\noaffiliation

\date{\today}

\begin{abstract}
Single-photon sources are subjected to a fundamental limitation in the speed of operation dictated by the spontaneous emission rate of quantum emitters (QEs). The current paradigm of the rate acceleration suggests coupling of a QE to a metal nanostructure, in particular, a metal nanoparticle (MNP). Here, we demonstrate that, in contrast to this approach, a MNP itself can behave as a quantum emitter. We determine both the first- and second-order correlation functions of light spontaneously emitted by a MNP strongly coupled to a QE and show that this light should exhibit sub-Poissonian photon statistics and perfect photon antibunching. This discovery opens a prospect to single-photon sources with unprecedented generation rates up to 100 THz.
\end{abstract}

% insert suggested PACS numbers in braces on next line
% insert suggested keywords - APS authors don't need to do this
%\keywords{}

%\maketitle must follow title, authors, abstract, \pacs, and \keywords
\maketitle

\begin{figure}
\includegraphics[width=\linewidth]{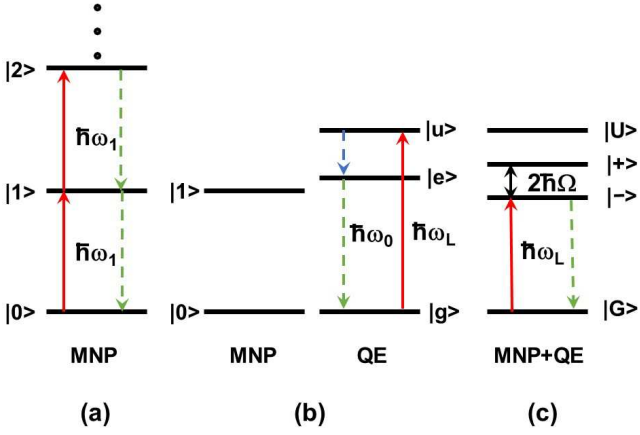}
\caption{\label{fig:levels} Energy levels of the system under consideration. (a) Plasmonic Fock states of an isolated MNP. The resonant laser excitation of frequency $\omega_L=\omega_1$ populates all Fock states $\mid N\rangle$ that leads to spontaneous emission at the transitions $\mid N\rangle\rightarrow\mid N-1\rangle$. (b) The levels of isolated MNP and QE involved in the coupling. (c) The levels of the compound system "MNP+QE". The laser excitation of frequency $\omega_L=\omega_-$ is resonant to the transition $\mid G\rangle\rightarrow\mid -\rangle$, but is out of resonance with the higher plasmonic Fock states.}
\end{figure}

\begin{figure}
\includegraphics[width=\linewidth]{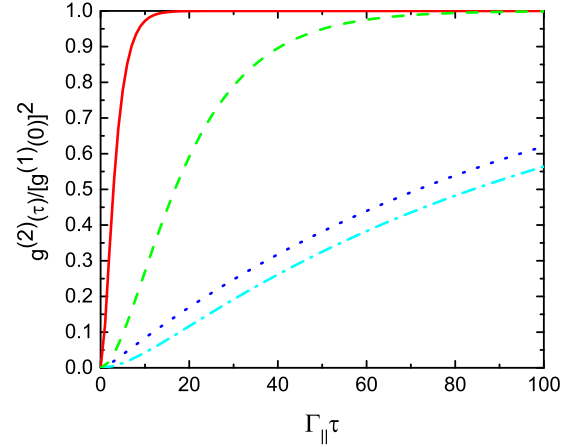}
\caption{\label{fig:g2} The normalized second-order correlation function of light emitted at the $\mid \pm\rangle\rightarrow\mid G\rangle$ transitions for non-resonant (NR) and resonant (R) excitation schemes and for $V_0/\Delta=1$, $(R+\gamma)/\Gamma_{\parallel}=0.01$ and $\Gamma_{\perp}/\Gamma_{\parallel}=0.5$. The nomenclature of the lines is as follows: R, $\mid -\rangle\rightarrow\mid G\rangle$ (red solid line); R, $\mid +\rangle\rightarrow\mid G\rangle$ (green dashed line); NR, $\mid -\rangle\rightarrow\mid G\rangle$ (blue dotted line); NR, $\mid +\rangle\rightarrow\mid G\rangle$ (cyan dash-dotted line). }
\end{figure}

% body of paper here - Use proper section commands
% References should be done using the \cite, \ref, and \label commands
{\it Introduction}. -- Recent progress in quantum technologies has opened the possibility to control the quantum properties of light and manipulate with single photons \cite{Lounis05,Eisaman11}. Single-photon sources have found numerous applications in spectroscopy, quantum optics and quantum technologies. They can be used for measurements of weak absorption, quantum information processing, quantum communication, quantum cryptography, quantum computing and quantum metrology \cite{Lounis05,Eisaman11,Dowling03,Brien07}. \\
A key element of a single-photon source is a quantum emitter (QE) such as atom, ion, molecule, color center, semiconductor nanocrystal or quantum dot. Single emitters being excited by an external source, either optically or electrically, emit light spontaneously delivering photons one at a time that is known as photon antibunching. To emit a next photon, the excitation-emission cycle should be completed that limits the repetition rate of single-photon emission and hence the speed of any single-photon device operation. The shortest emission lifetimes of 300 ps are typical for quantum dots which leads to the highest rate of photon emission of 1 GHz \cite{Lounis05}. \\
The current commonly adopted paradigm in accelerating the QE spontaneous emission suggests to introduce a dielectric or metallic (plasmonic) environment in the vicinity of the QE which modifies the electromagnetic local density of states for the optical transition \cite{Pelton15,Mortensen18}. The underlying mechanism of this modification is the Purcell effect \cite{Purcell} which can be alternatively considered as a result of the interaction of the QE with its own field reflected from the cavity walls or plasmonic nanostructure \cite{Barnes98}. The most pronounced effect takes place when the QE emission frequency is close to the frequencies of localized surface plasmons (LSPs) which are the normal modes of the collective electron oscillations in the nanostructure. The estimates show \cite{Khurgin16} that in such a case the plasmonic approach to the spontaneous emission enhancement can provide two orders of magnitude larger enhancement as compared to all-dielectric structures, reaching the ultimate level of single-photon emission of the order of 1 THz. Recently the experimental demonstrations of the ultrafast room-temperature single-photon emission with rate up to 80 GHz from quantum dots coupled to plasmonic nanostructures have been reported \cite{Hoang16,Wei19}.\\
In all studies up to date a plasmonic nanostructure in this context has been considered as a passive element which modifies the local environment of a QE. In the most simple configuration the nanostructure is reduced to a metal nanoparticle (MNP) acting as a nanoantenna which concentrates light from the far field and enhances the outcoupling of the QE radiation into the far field \cite{Maier11}. This phenomenon has been investigated in different configurations, both experimentally \cite{Hecht05,Anger06,Kuhn06} and theoretically \cite{Trugler08,Nerkararyan14,Delga14,Nerkararyan15,Torma15,Varguet16,Nerkararyan18}. One distinguishes between the weak coupling regime where only the relaxation dynamics of the QE is modified \cite{Nerkararyan14,Nerkararyan15} and the strong coupling regime where the emission spectrum is modified as well due to the fast Rabi oscillations between the QE and MNP states \cite{Trugler08,Delga14,Torma15,Varguet16,Lienau17,Nerkararyan18}. It has been also noticed that the photon antibunching in the QE emission can be controlled in the vicinity of a MNP \cite{Moradi18}.\\
In contrast to the situation discussed above, a MNP can emit light itself. Being optically excited above the onset of the interband transitions, noble-metal nanoparticles can demonstrate rather strong visible photoluminescence \cite{Wilcoxon98,Novotny03,Caruso04}. There is experimental evidence that the non-radiative relaxation channel in such a case can involve generation of LSPs which subsequently radiate light \cite{Caruso04}. The coupling of a MNP with a QE can lead to the enhancement of MNP photoluminescence \cite{Zhao18}. On the other hand, LSPs can be excited directly, either by fast electrons \cite{Crowell68,Yamamoto01} or by light resonant to the LSP mode \cite{Crowell68}. The emission rate of a dipole LSPs in a MNP is proportional to its volume and can reach the values of the order of 1 PHz = $10^3$ THz \cite{Crowell68,Bordo19}, three orders of magnitude larger than the ultimate rate expected for QEs coupled to metal nanostructures.\\
In the quantum description \cite{Bordo19}, the spectrum of the dominant dipole LSP mode is represented by an infinite equidistant ladder of the plasmonic Fock states $\mid N\rangle$ with $N$ being the number of quanta in the dipole plasmonic state [Fig. \ref{fig:levels} (a)]. Spontaneous emission in such a system results from downward transitions between the adjacent Fock states. Upon illumination by resonant laser light, which is described by a coherent state \cite{Lounis05}, all plasmonic Fock states are populated. There is therefore a certain probability that a few photons can be emitted simultaneously which is a disadvantage when one aims to achieve a single-photon emission.\\
This drawback can be avoided if a MNP is strongly coupled to a three-level QE and the latter one is excited by laser light through the transition which is non-resonant with the LSP mode [Fig. \ref{fig:levels} (b)] \cite{Note1}. Alternatively, the system can be excited resonantly to the LSP mode, but care should be taken to suppress the excitation of the higher plasmonic Fock states. \\
In the present Letter, we obtain the emission spectrum of the MNP strongly coupled to a QE and show that the emitted light is perfectly antibunched - the issue which, to the best of our knowledge, has so far not been discussed in the literature. We thus demonstrate that such a system can provide a platform for single-photon sources with unprecedented generation rates.\\
{\it Model.} -- We consider first the model system shown in Fig. \ref{fig:levels} (b). The QE, which has three energy levels $\mid g\rangle$, $\mid e\rangle$ and $\mid u\rangle$, is incoherently excited at the optically active transition $\mid g\rangle\rightarrow\mid u\rangle$. The transition frequency $\omega_0$ of the other optically active transition $\mid g\rangle\rightarrow\mid e\rangle$ is close to the frequency $\omega_1$ of the dipole LSP mode of the MNP. Let us note that such a model does not specify the MNP as far as its dipole LSP resonance can be engineered by the choice of the MNP composition, size, shape and its dielectric environment \cite{Lee06}. We assume also that the upper excited QE state $\mid u\rangle$ undergoes fast non-radiative decay to the state $\mid e\rangle$ and the frequency of the transition $\mid g\rangle\rightarrow\mid u\rangle$ is below the onset of interband transitions in the MNP, so that its photoluminescence is not excited. \\
The QE and the MNP are coupled with each other through the electrostatic interaction $\hat{V}$ so that the Hamiltonian of the system is written as $\hat{H}=\hat{H}_{QE}+\hat{H}_{MNP}+\hat{V}$, where $\hat{H}_{QE}$ and $\hat{H}_{MNP}$ are the Hamiltonians of the non-interacting components. It is reasonable to assume that the matrix elements of $\hat{V}$ are much less than the energy intervals $\hbar\omega_0$ and $\hbar\omega_1$, but may be comparable with $2\Delta=\hbar(\omega_0-\omega_1)$. Then in the lowest-order approximation the eigenstates of the system involved in the MNP-QE coupling are approximated by the states $\mid G\rangle=\mid g\rangle\mid 0\rangle$, $\mid U\rangle=\mid u\rangle\mid 0\rangle$ and
\begin{eqnarray}
\mid -\rangle = \cos\theta\mid g\rangle\mid 1\rangle -\sin\theta\mid e\rangle\mid 0\rangle,\label{eq:minus}\\
\mid +\rangle = \sin\theta\mid g\rangle\mid 1\rangle +\cos\theta\mid e\rangle\mid 0\rangle\label{eq:plus}
\end{eqnarray}
with $\tan 2\theta=V_0/\Delta$ and $V_0=\langle~1~\mid~\langle~g~\mid~\hat{V}~\mid~e~\rangle\mid~0~\rangle=V_0^*$. The states $\mid -\rangle$ and $\mid +\rangle$, Eqs. (\ref{eq:minus}) and (\ref{eq:plus}), can be identified as the QE states {\it dressed} by the MNP plasmonic field \cite{Meystre07}. The doublet of the dressed states is separated by the energy $2\hbar\Omega$ with $\Omega=\sqrt{\Delta^2+V_0^2}/\hbar$ being the frequency of the Rabi oscillations between the states $\mid g\rangle\mid 1\rangle$ and $\mid e\rangle\mid 0\rangle$ [Fig. \ref{fig:levels} (c)] \cite{Note3}. \\
{\it Relaxation processes}. -- In what follows, we will be interested in the correlation functions of light emitted at the transitions $\mid\pm\rangle\rightarrow\mid G\rangle$. The radiative relaxation rates are proportional to the square of the transition matrix elements of the dipole moment operator of the compound system "MNP+QE", $\hat{\bf D}^S=\hat{\bf D}+\hat{\bf d}$, where $\hat{\bf D}$ and $\hat{\bf d}$ are the dipole moment operators of the MNP and QE, respectively. For the transitions under discussion
\begin{eqnarray}
{\bf D}^S_{G-}={\bf D}_{01}\cos\theta-{\bf d}_{ge}\sin\theta\approx {\bf D}_{01}\cos\theta\equiv D_{01}^-, \label{eq:D1}\\
{\bf D}^S_{G+}={\bf D}_{01}\sin\theta+{\bf d}_{ge}\cos\theta\approx {\bf D}_{01}\sin\theta\equiv D_{01}^+, \label{eq:D2}
\end{eqnarray}
where we have taken into account that $d_{ge}\ll D_{01}$ and assumed $V_0\sim\mid\Delta\mid$ \cite{Note2}. Equations (\ref{eq:D1}) and (\ref{eq:D2}) imply that the contribution of the QE is negligible and the emission originates primarily from the MNP.\\
Under the same assumptions, this leads to the radiative relaxation rates $\Gamma_r^-\approx\Gamma_r\cos^2\theta$ and $\Gamma_r^+\approx\Gamma_r\sin^2\theta$ for the transitions $\mid-\rangle\rightarrow\mid G\rangle$ and $\mid+\rangle\rightarrow\mid G\rangle$, respectively, where $\Gamma_r$ is the radiative relaxation rate of an isolated MNP. In other words, the relaxation rate is determined by the squared amplitude of the admixture of the one-plasmon state.\\ 
Analogously, the non-radiative relaxation rates $\Gamma_{nr}^-\approx\Gamma_{nr}\cos^2\theta$ and $\Gamma_{nr}^+\approx\Gamma_{nr}\sin^2\theta$ due to the dominant contribution of the non-radiative decay rate of the MNP one-plasmon state, $\Gamma_{nr}$. We thus have for the total longitudinal (energy) relaxation rates $\Gamma_{\parallel }^-\approx\Gamma_{\parallel}\cos^2\theta$ and $\Gamma_{\parallel }^+\approx\Gamma_{\parallel}\sin^2\theta$ with $\Gamma_{\parallel}=\Gamma_{r}+\Gamma_{nr}$. By the same token, $\Gamma_{\perp }^-\approx\Gamma_{\perp}\cos^2\theta$ and $\Gamma_{\perp }^+\approx\Gamma_{\perp}\sin^2\theta$ for the transverse (phase) relaxation rates at the transitions $\mid\pm\rangle\rightarrow\mid G\rangle$ with $\Gamma_{\perp}$ being the phase relaxation rate of the plasmonic Fock state $\mid 1\rangle$ \cite{Bordo19}. Similar arguments can be applied to write $\gamma^{-}=\gamma\sin^2\theta$ and $\gamma^{+}=\gamma\cos^2\theta$ for the rates of the decays $\mid U\rangle\rightarrow\mid \pm\rangle$ with $\gamma$ being the non-radiative decay rate at the transition $\mid u\rangle\rightarrow\mid e\rangle$.\\
{\it Master equations.} -- The evolution of the system is described by the Liouville equation for the density matrix, $\rho$, which is split into the balance equations for the state populations
\begin{eqnarray}
\dot{\rho}_{GG}=-R\rho_{GG}+\Gamma_{\parallel}^-\rho_{--}+\Gamma_{\parallel}^+\rho_{++},\label{eq:balance1}\\
\dot{\rho}_{UU}=R\rho_{GG}-(\gamma^-+\gamma^+)\rho_{UU},\label{eq:balance2}\\
\dot{\rho}_{--}=\gamma^-\rho_{UU}-\Gamma_{\parallel}^-\rho_{--},\label{eq:balance3}\\
\dot{\rho}_{++}=\gamma^+\rho_{UU}-\Gamma_{\parallel}^+\rho_{++}\label{eq:balance4}
\end{eqnarray}
and the equations for the coherencies of interest
\begin{eqnarray}
\dot{\rho}_{-G}=-(\Gamma_{\perp}^-+i\omega_-)\rho_{-G}\\
\dot{\rho}_{+G}=-(\Gamma_{\perp}^++i\omega_+)\rho_{-G},
\end{eqnarray}
where $R$ is the pumping rate of the external field and $\omega_{\pm}=(\omega_0+\omega_1)/2\pm\Omega$ are the frequencies of the transitions $\mid\pm\rangle\rightarrow\mid  G\rangle$. Here we have neglected the rate of the decay $\mid U\rangle\rightarrow\mid G\rangle$ in comparison with the rates $\gamma^{\pm}$.\\
The set of Eqs. (\ref{eq:balance1})-(\ref{eq:balance4}) is solved by applying the Laplace transform with the initial condition $\rho_{GG}(0)=1$ and all other populations being equal to zero. In the steady-state limit ($t\gg \Gamma_{\parallel}^{-1}$) the populations of interest are given by
\begin{eqnarray}
\rho_{--}(\infty)=\frac{R\gamma^-\Gamma_{\parallel}^+}{Q}=\frac{R\gamma\Gamma_{\parallel}}{Q}\sin^4\theta,\label{eq:pop-}\\
\rho_{++}(\infty)=\frac{R\gamma^+\Gamma_{\parallel}^-}{Q}=\frac{R\gamma\Gamma_{\parallel}}{Q}\cos^4\theta,\label{eq:pop+}
\end{eqnarray}
where $Q=(\gamma^-+\gamma^++R)\Gamma_{\parallel}^-\Gamma_{\parallel}^++R(\gamma^-\Gamma_{\parallel}^++\gamma^+\Gamma_{\parallel}^-)$.\\
{\it Correlation functions of the emitted light.} -- The electromagnetic field spontaneously emitted by a compound quantum dipole source in the far-field region is written as a sum of the positive-frequency and negative-frequency parts \cite{Swain80}
\begin{equation}
{\bf E}({\bf r},t)={\bf E}^+({\bf r},t)+{\bf E}^-({\bf r},t),
\end{equation}
where
\begin{equation}
{\bf E}^+({\bf r},t)=-\frac{\omega_{ul}^2}{4\pi\epsilon_0c^2r^3}\left[\left({\bf D}^S_{lu}\times{\bf r}\right)\times{\bf r}\right]\sigma_-(\hat{t})
\end{equation}
and ${\bf E}^-({\bf r},t)=\left[{\bf E}^+({\bf r},t)\right]^{\dagger}$. Here $\omega_{ul}$ is the frequency of the transition from the upper state $\mid u\rangle$ to the lower state $\mid l\rangle$, ${\bf D}^S_{lu}$ is the transition dipole moment of the system, $\sigma_-=\mid l\rangle\langle u\mid$ is the lowering operator, $c$ is the speed of light in vacuum and the time argument $\hat{t}\equiv t-r/c$ takes into account retardation.\\
The experimentally observed quantities can be expressed in terms of the first- and second-order correlation functions \cite{Swain80}
\begin{eqnarray}
\langle E^-(t)E^+(t+\tau )\rangle\propto g^{(1)}(t,t+\tau )\nonumber\\
\equiv \text{Tr}_{S+R}\left[\rho_{S+R}(0)\sigma_+(\hat{t})\sigma_-(\hat{t}+\tau )\right] \label{eq:g1}
\end{eqnarray}
and
\begin{eqnarray}
\langle E^-(t)E^-(t+\tau)E^+(t+\tau )E^+(t)\rangle\propto g^{(2)}(t,t+\tau )\nonumber\\
\equiv \text{Tr}_{S+R}\left[\rho_{S+R}(0)\sigma_+(\hat{t})\sigma_+(\hat{t}+\tau )\sigma_-(\hat{t}+\tau )\sigma_-(\hat{t})\right],\label{eq:g2}\nonumber\\
\end{eqnarray}
where $\sigma_+=\mid u\rangle\langle l\mid$ is the raising operator and the averaging is taken over the "system $(S)$ + reservoir $(R)$" states with the reservoir being the electromagnetic vacuum $\mid\emptyset\rangle$, so that $\rho_{S+R}(t)=\rho(t)\cdot\mid\emptyset\rangle\langle\emptyset\mid$.\\
The correlation functions $g^{(1)}(t,t+\tau)$ and $g^{(2)}(t,t+\tau)$ are found with the use of the quantum regression theorem \cite{Lax68,Mollow69,Swain80} which relates the $\tau$-evolution of the operators under the trace symbol in Eqs. (\ref{eq:g1}) and (\ref{eq:g2}) with the evolution of $\rho(\tau)$. As a result, one obtains for the steady-state limits $g_{\pm}^{(1)}(\tau)\equiv \lim_{t\to\infty}g_{\pm}^{(1)}(t,t+\tau)$ (here and in what follows the subscript or superscript "$\pm$" indicates the transitions $\mid \pm\rangle\rightarrow\mid G\rangle$)
\begin{equation}
g_{\pm}^{(1)}(\tau)=\rho_{\pm\pm}(\infty)e^{-(\Gamma_{\perp}^{\pm}+i\omega_{\pm})\tau}\label{eq:g1tau},
\end{equation}
where the steady-state populations $\rho_{\pm\pm}(\infty)$ are given by Eqs. (\ref{eq:pop-}) and (\ref{eq:pop+}). Accordingly, the spectra of spontaneous emission at these transitions are found as 
\begin{figure}
\includegraphics[width=\linewidth]{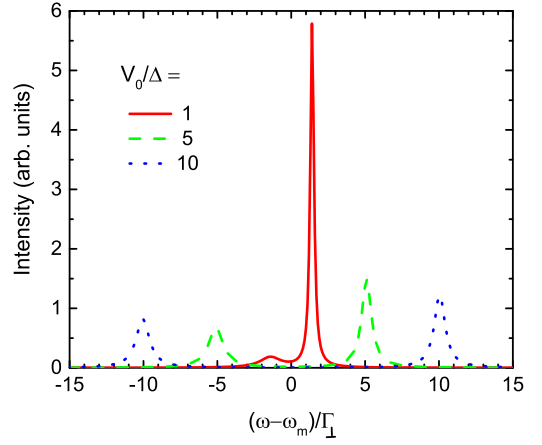}
\caption{\label{fig:spectrum} The emission spectrum of the MNP coupled to a QE for $\Delta/(\hbar \Gamma_{\perp})=1$ and different coupling strengths shown in the inset. $\omega_m\equiv (\omega_0+\omega_1)/2$. }
\end{figure}
\begin{equation}
S_{\pm}(\omega)=\int_{-\infty}^{\infty}e^{i\omega\tau}g_{\pm}^{(1)}(\tau)d\tau=\frac{2\Gamma_{\perp}^{\pm}\rho_{\pm\pm}(\infty)}{(\omega-\omega_{\pm})^2+(\Gamma_{\perp}^{\pm})^2},
\end{equation}
where the functions $g_{\pm}^{(1)}(\tau)$ for negative $\tau$ are defined as $g_{\pm}^{(1)}(-\tau)=g_{\pm}^{(1)*}(\tau)$ \cite{Mollow69}. The linewidths of the doublet are related with each other as $\Gamma_{\perp}^+/\Gamma_{\perp}^-=\tan^2\theta$ while their intensities at the maximums, taking into account Eqs. (\ref{eq:D1}) and (\ref{eq:D2}), are related as $I_+/I_-=\cot^4\theta$ (see Fig. \ref{fig:spectrum}).\\
The steady-state limits of the second-order coherences, $g_{\pm}^{(2)}(\tau)\equiv \lim_{t\to\infty}g_{\pm}^{(2)}(t,t+\tau)$, are given by
\begin{eqnarray}
g_{\pm}^{(2)}(\tau)\approx \left[g_{\pm}^{(1)}(0)\right]^2\nonumber\\
\times \left(1-\frac{R+\gamma}{\Gamma_{\parallel}^{\pm}}-e^{-(R+\gamma)\tau}+\frac{R+\gamma}{\Gamma_{\parallel}^{\pm}}e^{-\Gamma_{\parallel}^{\pm}\tau}\right),\label{eq:g2tau}
\end{eqnarray}
where we have adopted a reasonable assumption that $\gamma,R\ll \Gamma_{\parallel}$ and have taken the lowest order in $\gamma/\Gamma_{\parallel}$ and $R/\Gamma_{\parallel}$. We have also used here the relation $\rho_{\pm\pm}(\infty)=g_{\pm}^{(1)}(0)$ [see Eq. (\ref{eq:g1tau})]. \\
Equation (\ref{eq:g2tau}) reveals that $g_-^{(2)}(0)=g_+^{(2)}(0)=0$, i.e. both lines of the doublet exhibit perfect photon antibunching. When the delay between two photons $\tau\gg (R+\gamma)^{-1}$, $g_{\pm}^{(2)}(\tau)\approx \left[g_{\pm}^{(1)}(0)\right]^2$ and two fields are completely uncorrelated. For finite delays $\tau > 0$ $g_{\pm}^{(2)}(\tau)<\left[g_{\pm}^{(1)}(0)\right]^2$ that indicates the sub-Poissonian photon statistics (Fig. \ref{fig:g2}) \cite{Stevens13}.\\
{\it Photon generation rate.} -- The probability of photon emission per unit time, $p$, is determined as the product of the probability that the system occupies the excited state with the radiative emission rate from this state, i.e. $p_{\pm}=\rho_{\pm\pm}(\infty)\Gamma_r^{\pm}$, where we assume the steady-state operation. The analysis of the expressions for the dressed state populations, Eqs. (\ref{eq:pop-}) and (\ref{eq:pop+}), reveals that, not depending on the ratio between the rate constants $R$ and $\gamma$, $p_{\pm}\sim \eta\min(R,\gamma)$, where $\eta=\Gamma_r/\Gamma_{\parallel}$ is the quantum yield of the MNP light emission. \\
The emission quantum yield for Au nanorods of diameter 20 nm and length 60 nm is about 10 \% \cite{Mulvaney02}, while that for Ag spheres of radius 20 nm is about 30 \% \cite{Crowell68}. One can expect, however, that $\eta$ increases with the MNP size. For a spherical MNP of radius $R$ the radiative relaxation rate scales as $R^3$ while the size-dependent part of the non-radiative damping rate scales as $1/R$ \cite{Kreibig95}. \\
The non-radiative relaxation rates in quantum dots can be of the order of 10$^{12}$ s$^{-1}$ \cite{Egeler92} that can provide the single-photon emission rate up to 1 THz. Although this value is far above the generation rates of the currently available single-photon sources, this stage in the excitation channel presents a "bottleneck" as compared to the ultimate radiative relaxation rates achievable in MNPs.\\
{\it Alternative scenario.} -- An alternative approach could be a direct excitation of the dressed states by a laser light of frequency $\omega_L$ close to either $\omega_-$ or $\omega_+$ far below saturation. Due to the strong MNP-QE coupling the two transitions $\mid G\rangle\rightarrow\mid\pm\rangle$ are well spectrally resolved and one can excite either the $\mid -\rangle$ or the $\mid +\rangle$ state, not exciting the higher plasmonic Fock states [Fig. \ref{fig:levels} (c)]. In such a case, the correlation functions of the spontaneously emitted light can be found from the solution of the optical Bloch equations in the lowest non-vanishing order with respect to the Rabi frequency of the corresponding transition, $\Omega_L^{(\pm)}=D_{01}^{\pm}E_L/\hbar$ with $E_L$ being the laser field \cite{Bordo19}. In the particularly simple case of exact resonance ($\omega_L=\omega_{\pm}$) this approach gives
\begin{eqnarray}
g_{\pm}^{(2)}(\tau)\approx \left[g_{\pm}^{(1)}(0)\right]^2\nonumber\\
\times \left(1-\frac{\Gamma_{\perp}^{\pm}}{\Gamma_{\perp}^{\pm}-\Gamma_{\parallel}^{\pm}}e^{-\Gamma_{\parallel}^{\pm}\tau}+\frac{\Gamma_{\parallel}^{\pm}}{\Gamma_{\perp}^{\pm}-\Gamma_{\parallel}^{\pm}}e^{-\Gamma_{\perp}^{\pm}\tau}\right),
\end{eqnarray}
that reveals both the sub-Poissonian photon statistics and perfect antibunching as before (Fig. \ref{fig:g2}).\\
The corresponding steady-state populations of the dressed states equal $\rho_{\pm\pm}(\infty)=2\Omega_L^{(\pm)2}/(\Gamma_{\perp}^{\pm}\Gamma_{\parallel}^{\pm})$. Then if one chooses, for example, $\Omega_L^{(\pm)2}/(\Gamma_{\perp}^{\pm}\Gamma_{\parallel}^{\pm})\sim 0.05$, the single-photon generation rate can reach the value $\sim 0.1\times\Gamma_r^{\pm}\sim 10^2$ THz.\\
{\it Conclusion.} -- In this Letter, we have demonstrated that a MNP strongly coupled to a QE should spontaneously emit light which obeys sub-Poissonian statistics and exhibit perfect antibunching. We have considered two different scenarios of excitation: (i) the MNP is excited non-resonantly via coupling with a three-level QE, and (ii) the MNP-QE compound is excited resonantly far below saturation. We have found that in the first case the single-photon generation rate is fundamentally limited by the non-radiative relaxation rate in the excitation channel. On the contrary, the latter scenario is seen as a simple and promising approach which can ensure unprecedented repetition rate of single-photon emission of the order of 100 THz.\\
To realize the full potential of a MNP as a single-photon source, it should be coupled to a high-Q resonant cavity which can enhance the spontaneous emission, channel the emitted photons and narrow the spectral range of emission \cite{Lounis05}. This can be implemented, for example, in a photonic nanowire which has been demonstrated to be a highly efficient platform for a single-photon source \cite{Claudon10}.


\begin{thebibliography}{99}
\bibitem{Lounis05} B.~Lounis and M.~Orrit, Rep. Prog. Phys. {\bf 68}, 1129 (2005).
\bibitem{Eisaman11} M.D.~Eisaman, J.~Fan, A.~Migdal, and S.V.~Polyakov, Rev. Sci. Instrum. {\bf 82}, 071101 (2011).
\bibitem{Dowling03} J.P.~Dowling and G.J.~Milburn, Philos. Trans. R. Soc. London, Ser. A {\bf 361}, 1655 (2003). 
\bibitem{Brien07} J.L.~O'Brien, Science {\bf 318}, 1567 (2007).
\bibitem{Pelton15} M.~Pelton, Nat. Photonics {\bf 9}, 427 (2015).
\bibitem{Mortensen18} A.I.~Fern{\'a}ndez-Dom{\'i}nguez, S.I.~Bozhevolnyi, and N.A.~Mortensen, ACS Photonics {\bf 5}, 3447 (2018).
\bibitem{Purcell} E.M.~Purcell, Phys. Rev. {\bf 69}, 681 (1946).
\bibitem{Barnes98} W.L.~Barnes, J. Mod. Opt. {\bf 45}, 661 (1998).
\bibitem{Khurgin16} S.I.~Bozhevolnyi and J.B.~Khurgin, Optica {\bf 3}, 1418 (2016).
\bibitem{Hoang16} T.B.~Hoang, G.M.~Akselrod, and M.H.~Mikkelsen, Nano Lett. {\bf 16}, 270 (2016).
\bibitem{Wei19} W.~Wei, X.~Yan, J.~Liu, B.~Shen, W.~Luo, X.~Ma, and X.~Zhang, Nanomater. {\bf 9}, 671 (2019).
\bibitem{Maier11} V.~Giannini, A.I.~Fern{\'a}ndez-Dom{\'i}nguez, S.C.~Heck, and S.A.~Maier, Chem. Rev. {\bf 111}, 3888 (2011).
\bibitem{Hecht05} J.N.~Farahani, D.W.~Pohl, H.-J.~Eisler, and B.~Hecht, Phys. Rev. Lett. {\bf 95}, 017402 (2005).
\bibitem{Anger06} P.~Anger, P.~Bharadwaj, and L.~Novotny, Phys. Rev. Lett. {\bf 96}, 113002 (2006).
\bibitem{Kuhn06} S.~K{\"u}hn, U.~H{\aa}kanson, L.~Rogobete, and V.~Sandoghdar, Phys. Rev. Lett. {\bf 97}, 017402 (2006).
\bibitem{Trugler08} A.~Tr{\"u}gler and U.~Hohenester, Phys. Rev. B {\bf 77}, 115403 (2008).
\bibitem{Nerkararyan14} K.V.~Nerkararyan and S.I.~Bozhevolnyi, Opt. Lett. {\bf 39}, 1617 (2014).
\bibitem{Delga14} A.~Delga, J.~Feist, J.~Bravo-Abad, and F.J.~Garcia-Vidal, Phys. Rev. Lett. {\bf 112}, 253601 (2014).
\bibitem{Nerkararyan15} K.V.~Nerkararyan and S.I.~Bozhevolnyi, Faraday Discuss. {\bf 178}, 295 (2015).
\bibitem{Varguet16} H.~Varguet, B.~Rousseaux, D.~Dzsotjan, H.R.~Jauslin, S.~Gu{\'e}rin, and G. Colas des Francs, Opt. Lett. {\bf 41}, 4480 (2016).
\bibitem{Nerkararyan18} K.V.~Nerkararyan, T.S.~Yezekyan, and S.I.~Bozhevolnyi, Phys. Rev. B {\bf 97}, 045401 (2018).
\bibitem{Torma15} P.~T{\"o}rm{\"a} and W.L.~Barnes, Rep. Prog. Phys. {\bf 78}, 013901 (2015), and references therein.
\bibitem{Lienau17} P.~Vasa and C.~Lienau, ACS Photonics {\bf 5}, 2 (2018).
\bibitem{Moradi18} T.~Moradi, M.~Bagheri Harouni, and M.H.~Naderi, Sci. Rep. {\bf 8}, 12435 (2018).
\bibitem{Wilcoxon98} J.P.~Wilcoxon, J.E.~Martin, F.~Parsapour, B.~Wiedenman, and D.F.~Kelley, J. Chem. Phys. {\bf 108}, 9137 (1998).
\bibitem{Novotny03} M.R.~Beversluis, A.~Bouhelier, and L.~Novotny, Phys. Rev. B {\bf 68}, 115433 (2003).
\bibitem{Caruso04} E.~Dulkeith, T.~Niedereichholz, T.A.~Klar, J.~Feldmann, G.~von~Plessen, D.I.~Gittins, K.S.~Mayya, and F.~Caruso, Phys. Rev. B {\bf 70}, 205424 (2004).
\bibitem{Zhao18} J.~Zhao, Y.~Cheng, H.~Shen, Y.Y.~Hui, T.~Wen, H.-C.~Chang, Q.~Gong, and G.~Lu, Sci. Rep. {\bf 8}, 3605 (2018).
\bibitem{Crowell68} J.~Crowell and R.H.~Ritchie, Phys. Rev. {\bf 172}, 436 (1968).
\bibitem{Yamamoto01} N.~Yamamoto, K.~Araya, and F.J.~Garc{\'i}a de Abajo, Phys. Rev. B {\bf 64}, 205419 (2001).
\bibitem{Bordo19} V.G.~Bordo, J. Opt. Soc. Am. B {\bf 36}, 323 (2019).
\bibitem{Note1} Let us note that such a configuration has been considered in Refs. \cite{Trugler08,Nerkararyan14,Nerkararyan15} to describe the QE-MNP coupling.
\bibitem{Lee06} K.-S. Lee and M.A. El-Sayed, J. Phys. Chem. B {\bf 110}, 19220 (2006).
\bibitem{Meystre07} P.~Meystre and M.~Sargent III, {\it Elements of Quantum Optics} (Springer-Verlag, Berlin, 2007).
\bibitem{Note3} Let us note that the notation of the dressed states implies that the state $\mid -\rangle$ is the lower state for positive $\Delta$ and the upper state for negative $\Delta$.
\bibitem{Note2} For a spherical MNP of radius $R$ in vacuum the transition dipole moment is given by $D_{01} = \sqrt{\hbar\omega_1R^3/2}$ and can be of the order of $10^3$ D (Debye); see Ref. \cite{Bordo19} for detail. In such a case if $d_{ge}\sim 1$ D, the approximations in Eqs. (\ref{eq:D1}) and (\ref{eq:D2}) are valid if $\mid\tan\theta\mid,\mid\cot\theta\mid\ll 10^3$.
\bibitem{Swain80} S.~Swain, in {\it Advances in Atomic and Molecular Physics}, edited by D.R.~Bates and B.~Bederson (Academic, New York, 1980), Vol. 16, p. 159.
\bibitem{Lax68} M.~Lax, Phys. Rev. {\bf 172}, 350 (1968).
\bibitem{Mollow69} B.R.~Mollow, Phys. Rev. {\bf 188}, 1969 (1969).
\bibitem{Stevens13} M.J.~Stevens, in {\it Single-Photon Generation and Detection}, edited by A.~Migdall, S.~Polyakov, J.~Fan, and J.~Bienfang (Academic, New York, 2013), Vol. 45, p. 25.
\bibitem{Mulvaney02} C.~S{\"o}nnichsen, T.~Franzl, T.~Wilk, G.~von Plessen, J.~Feldmann, O.~Wilson, and P.~Mulvaney, Phys. Rev. Lett. {\bf 88}, 077402 (2002).
\bibitem{Kreibig95} U.~Kreibig and M.~Vollmer, {\it Optical Properties of Metal Clusters} (Springer-Verlag, Berlin, 1995).
\bibitem{Egeler92} U.~Bockelmann and T.~Egeler, Phys. Rev. B {\bf 46}, 15574 (1992).
\bibitem{Claudon10} J.~Claudon, J.~Bleuse, N.S.~Malik, M.~Bazin, P.~Jaffrennou, N.~Gregersen, C.~Sauvan, P.~Lalanne, and J.-M.~G{\'e}rard, Nat. Photonics {\bf 4}, 174 (2010).



\end{thebibliography}
\end{document}